\documentclass[prl,aps,twocolumn,preprintnumbers, showpacs, nofootinbib,superscriptaddress,notitlepage]{revtex4-1}
\usepackage{mathrsfs}
\usepackage{amsfonts}
\usepackage{amsmath}
\usepackage{slashed}
\usepackage{array}
\usepackage{verbatim}
\usepackage{epsfig}
\usepackage{graphicx}
\usepackage{color}
\usepackage{hyperref}
\usepackage{footmisc}

\usepackage[dvipsnames]{xcolor}
\usepackage{subfigure}
\usepackage{float}
\usepackage{amssymb}
\usepackage{pifont}
\usepackage{setspace}
\usepackage{multirow}
\usepackage{booktabs}
\usepackage[normalem]{ulem}
\usepackage{placeins}

\newcommand{\beq}{\begin{eqnarray}}
\newcommand{\eeq}{\end{eqnarray}}

\begin{document}
\title{Computability of GPDs near $x=\pm\xi$ in Lattice QCD}

\author{Yushan Su}
\email{ysu12345@umd.edu}
\affiliation{Department of Physics, University of Maryland, College Park, MD 20742, USA}

\author{Xiangdong Ji}
\email{Corresponding author: xdji@sjtu.edu.cn}
\affiliation{Tsung-Dao Lee Institute and School of Physics and Astronomy, Shanghai Jiao Tong University, Shanghai 201210, China}

\author{Yizhuang Liu}
\email{yizhuang.liu@uj.edu.pl}
\affiliation{Institute of Theoretical Physics,
Jagiellonian University, 30-348 Kraków, Poland}

\author{Rui Zhang}
\email{Corresponding author: rzhang93@mit.edu}
\affiliation{Center for Theoretical Physics - a Leinweber Institute, Massachusetts Institute of Technology, Cambridge, MA 02139, USA}

\preprint{MIT-CTP/6078}
\begin{abstract}
In lattice QCD computations of generalized parton distributions (GPDs), 
the large momentum expansion generally requires all hard scales, 
$2|x\pm\xi|P^z$ and $2|1\pm x|P^z$, to be much larger than 
$\Lambda_{\rm QCD}$. We show that this condition can be relaxed for 
$2|x\pm\xi|P^z$ at large $\xi$, making the important $x\sim\pm\xi$ regions accessible to lattice calculations and considerably expanding the region of computability. Revisiting previous lattice results with 
complete one-loop matching, we obtain the expected partonic threshold 
behavior---GPDs continuous at $x=\pm\xi$ but with discontinuous 
derivatives---which has not previously been observed on the lattice. 
We thus obtain, for the first time, important prediction for GPDs in the 
distribution-amplitude-like region, which smoothly connects the quark and 
antiquark PDF-like behaviors.

\end{abstract}

\maketitle

\textit{Introduction:}
Generalized parton distributions (GPDs)~\cite{Muller:1994ses,Ji:1996ek,Ji:2016djn} describe correlated quark and gluon distributions in longitudinal momentum and transverse space, providing rich information on the spin, mass, and force distributions inside the nucleon~\cite{Ji:1996ek,Ji:2021mtz,Polyakov:2018zvc}. Experimental studies of GPDs are major scientific goals at Jefferson Lab~\cite{Dudek:2012vr} and future Electron-Ion Colliders (EICs)~\cite{Accardi:2012qut,AbdulKhalek:2021gbh,Anderle:2021wcy}. GPDs can be constrained through hard exclusive processes such as deeply virtual Compton scattering (DVCS)~\cite{Ji:1996ek,Ji:1996nm} and deeply virtual meson production (DVMP)~\cite{Radyushkin:1996ru,Collins:1996fb,Mankiewicz:1997uy}. More recently, GPDs have also been extracted through global analyses combining experimental and lattice inputs within the universal moment parameterization (GUMP) framework~\cite{Guo:2022upw,Guo:2023ahv,Zhang:2024djl,Guo:2024wxy,Guo:2025muf}.

Considerable effort has been devoted to computing GPDs and their moments in lattice quantum chromodynamics (QCD)~\cite{Cichy:2026xrh}. More recently, many studies~\cite{Ji:2015qla,Xiong:2015nua,Chen:2019lcm,Alexandrou:2019dax,Liu:2019urm,Lin:2020rxa,Alexandrou:2020zbe,Bhattacharya:2020xlt,Bhattacharya:2020jfj,Alexandrou:2021bbo,Lin:2021brq,Scapellato:2022mai,Bhattacharya:2022aob,Ma:2022ggj,Ma:2022gty,Yao:2022vtp,Bhattacharya:2023nmv,Bhattacharya:2023jsc,Braun:2023alc,Lin:2023gxz,Holligan:2023jqh,Ding:2024saz,Braun:2024snf,Chu:2025jsi,Holligan:2025baj,Chu:2025kew,Bhattacharya:2025yba} have directly computed the $x$ dependence of GPDs or analyzed relevant theoretical questions using the large momentum expansion (LaMET)~\cite{Ji:2013dva,Ji:2014gla,Ji:2020byp,Ji:2022ezo,Ji:2024oka}. These pioneering calculations have significantly advanced our understanding of GPDs, particularly their $x$ dependence. An important question remains, however, as to whether GPDs are calculable within LaMET near $x=\pm\xi$, where the longitudinal momentum of one of the partons approaches zero. Previous studies have suggested potential difficulties in this region. In the threshold-resummation framework of Ref.~\cite{Holligan:2025baj}, the scale $2|x\pm\xi|P^z$ enters a perturbative hard kernel. As $x$ approaches $\mp\xi$, this scale becomes of order $\Lambda_{\rm QCD}$, apparently invalidating the perturbative expansion. Moreover, several lattice calculations of GPDs at nonzero skewness~\cite{Alexandrou:2020zbe,Alexandrou:2021bbo,Holligan:2023jqh,Chu:2025kew} exhibit discontinuities at $x=\pm\xi$ after perturbative matching.

Clarifying the computability of GPDs near $x=\pm\xi$ is therefore crucial. 
If the standard effective-field-theory (EFT) argument applies, requiring
$2|x\pm\xi|P^z\gg\Lambda_{\rm QCD}$, then for the currently accessible 
momenta $P^z\sim 2$--$3$ GeV, one would require roughly 
$|x\pm\xi|\gtrsim 0.2$. This condition would severely limit the predictive 
power of the LaMET expansion. Indeed, given the presently accessible range 
$-0.8<x<0.8$, excluding windows of width $\sim0.4$ around $x=\pm\xi$ would 
leave little kinematic region in which LaMET calculations are reliable.

Fortunately, we find that the standard EFT requirement $2|x\pm\xi|P^z\gg\Lambda_{\rm QCD}$ is unnecessary. In fact, GPDs remain calculable near $x=\pm\xi$, much as DVCS can probe GPDs at $x=\pm\xi$ without encountering an infrared problem~\cite{Ji:1996ek,Ji:1998xh, Radyushkin:1997ki,Collins:1998be}. In this paper, we present theoretical arguments showing that potential soft contributions near $x=\pm\xi$ are power suppressed in the large momentum expansion. This is analogous to DVCS, where soft physics associated with the final-state real photon contributes only at subleading power. Furthermore, we reanalyze lattice QCD data at nonzero skewness~\cite{Alexandrou:2020zbe, Alexandrou:2021bbo,Holligan:2023jqh,Chu:2025kew} with a careful implementation of the perturbative matching and find that GPDs are continuous at $x=\pm\xi$, while their derivatives are not. These results provide strong evidence that GPDs near $x=\pm\xi$ are calculable within LaMET. Consequently, LaMET can offer powerful predictions for GPDs in the distribution-amplitude-like region, where little information is currently available from other resources.
\begin{figure*}[htbp]
    \centering
    \includegraphics[width=0.3\linewidth]{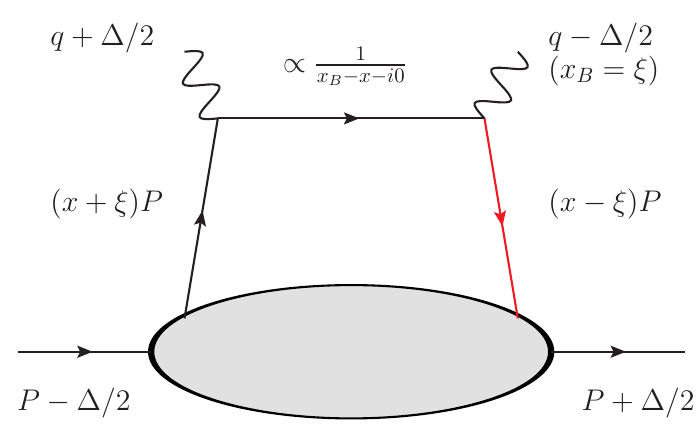} 
    \quad \quad \quad 
    \includegraphics[width=0.3\linewidth]{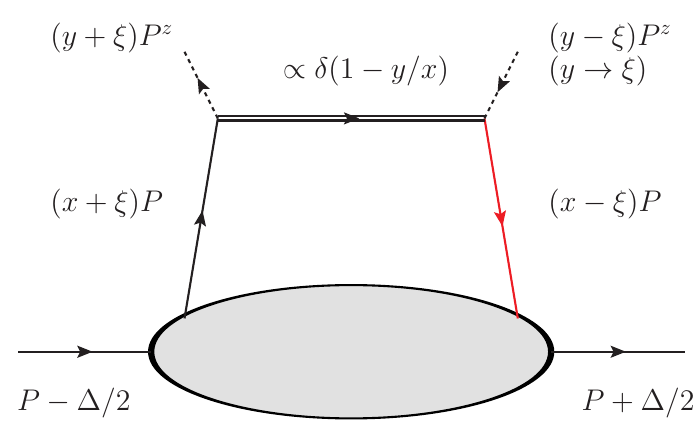}
    \caption{The handbag diagrams for DVCS (left panel) and quasi-GPD (right panel). In the DVCS case, the final-state real photon restricts $x_B=\xi$. The leading order imaginary amplitude mostly comes from $x=\xi$, which probes GPD at $x = \xi$. Similarly for the quasi-GPD, the tree-level diagram is proportional to $\delta(1-y/x)$, and the $y \rightarrow \xi$ limit directly measures GPD at $x = \xi$. }
    \label{fig:handbag}
\end{figure*}

\vspace{0.1in}
\textit{LaMET's predictive power for $x \sim \pm \xi$:}
To understand why GPDs at $ x=\pm \xi$ are computable as long as $P^z$ is perturbative, one may review a similar situation of deeply-virtual Compton scattering (DVCS)~\cite{Ji:1996ek,Ji:1996nm}. In this process, an incoming hadron with momentum $P-\Delta/2$ absorbs a virtual photon of momentum $q+\Delta/2$ and then emits a real photon with momentum $q-\Delta/2$ while recoiling with momentum $P+\Delta/2$. The Compton amplitude takes the form,
\begin{align}
T^{\mu \nu} = i \int d^4 z e^{i q \cdot z}\left\langle P+\frac{\Delta}{2}\right| T J^\nu\left(\frac{z}{2}\right) J^\mu\left(-\frac{z}{2}\right)\left|P-\frac{\Delta}{2}\right\rangle  \ ,
\end{align}
where $J^{\mu}$ is the electromagnetic current. $T^{\mu\nu}$ is a function of virtuality $Q^2 \equiv -q^2$, Bjorken parameter $x_B \equiv Q^2/(2P\cdot q)$, skewness $\xi \equiv -\Delta^{+}/(2P^{+})$, and Lorentz invariant momentum transfer $t \equiv \Delta^2$. Here, the light-cone coordinate is defined as $P^{\pm} \equiv (P^t \pm P^z)/\sqrt{2}$. In the Bjorken limit, namely $Q^2 \rightarrow \infty$ with fixed $x_B=\xi$ (real photon condition), and $\Delta^2$, the Compton amplitude can be factorized into GPD $ F(x,\xi,t)$ convoluted with the coefficient function $\hat{C}$~\cite{Ji:1998xh}, 
\begin{align}\label{eq:DVCSfac}
T^{ij} = -g^{ij} \int_{-1}^{1} \frac{dx}{x} \hat{C}\left(\frac{\xi}{x},\frac{\mu}{Q}\right) F(x,\xi,t) + ...
\end{align}
where the sum over quark flavors, the gluon contribution, and the asymmetric parts are omitted for simplicity.  The real photon emission does not generate problems for factorization~\cite{Ji:1998xh,Collins:1998be,Radyushkin:1997ki} as some had initially suspected. This is because the non-perturbative physics associated with the (on-shell) photon structure is sub-leading in power $1/Q^2$~\cite{Ji:1998xh}. 

 The leading handbag diagram for the Compton amplitude can be found in the left panel of Fig.~\ref{fig:handbag}, where its crossing is not presented for convenience.  The two partons carry momenta $(x+\xi)P$ and $(x-\xi)P$, respectively. when $x_B = \xi$, the leading imaginary contribution comes from the on-shell intermediate quark with $x = \pm x_B$, which probes $F(x=\pm\xi, \xi, t)$. Therefore, the experimental scattering cross-section can be used to measure the GPD at the transition points $x = \pm \xi$ directly, where one of the quarks is apparently soft (zero-mode). 

One may also understand the DVCS factorization by starting with the final-state photon as being highly virtual, with invariant mass $Q^2 (\xi-x_B)/x_B$~\cite{Ji:1998xh}. In the soft 
photon limit, soft-collinear modes can produce double logs in $\xi-x_B$~\cite{Schoenleber:2022myb}. For example, the one-loop coefficient function in Eq.~(34) of Ref.~\cite{Ji:1998xh} contains the following term,
\begin{align}
&-\frac{x}{\xi}\left[\frac{\left(2 \xi  x_B+x_B^2+2 \xi^2-x^2\right)}{\left(x_B^2-x^2\right) \left(x^2-\xi^2\right)}\left(3-\ln \left(\frac{x_B-\xi}{x_B}\right)\right) \right.\nonumber\\
&\left. \quad\quad\quad  -3 \frac{1}{x^2-\xi^2}\right] \left(x_B-\xi \right) \ln \left(\frac{x_B-\xi}{x_B}\right)    \ .
\end{align}
It appears that summing these double logs generates an infrared scale $Q^2 (\xi-x_B)/x_B$ in the QCD coupling. However, there is a linear factor $(x_B-\xi)$ that multiplies these double logs, power-suppressing them when $Q^2(\xi-x_B)/x_B$ approaches $\Lambda_{\rm QCD}^2$. This is consistent with that in the real photon limit  $x_B \rightarrow \xi$, photon distribution amplitude (DA) brings in a power-suppressed contribution $O\left(\Lambda_{\rm QCD}^2/Q^2\right)$.

\begin{table*}[htbp]
    \centering
    \begin{tabular}{|c|c|c|c|c|}
      \hline
      Works & ETMC20~\cite{Alexandrou:2020zbe} & ETMC21~\cite{Alexandrou:2021bbo} & MSULat23~\cite{Holligan:2023jqh} &  ETMC25~\cite{Chu:2025kew} \\ \hline 
      GPDs & $H,E,\tilde{H},\tilde{E}$ & $H_T,E_T,\tilde{H}_T,\tilde{E}_T$ & $H,E$ & $H,E$ \\ \hline 
      $\xi$ & $\pm$1/3 & $\pm$1/3 & 0.1 & 1/7,$\pm$1/5,$\pm$1/2 \\ \hline
      $a$ (fm) & 0.093 & 0.093 & 0.09 & 0.093 \\
      \hline
      $m_{\pi}$ (MeV) & 260 & 260 & 130 & 260 \\
      \hline
      $P^z$ (GeV) & 1.25 & 1.25 & 2.22 & 1.46, 1.04, 0.83 \\ 
      \hline
    \end{tabular}
    \caption{Lattice QCD datasets for the non-zero skewness nucleon GPDs. Their information, including GPD types, skewness $\xi$, lattice spacing $a$, pion mass $m_{\pi}$, and hadron momentum $P^z$, is listed in the table. }
    \label{tab:setlatdat}
\end{table*}

The situation with computing GPDs starting from quasi-GPDs are exactly analogous, allowing $x=\pm\xi$ regions calculable without introducing non-perturbative physics 
 through infrared scale $(x\pm \xi)P^z$. 
The quasi-GPD can be defined as~\cite{Ji:2013dva,Ji:2015qla},
\begin{align}\label{eq:qgpd}
&\tilde{F}(y,\xi,t,P^z) \equiv P^{z}\int \frac{d z}{2\pi} e^{-i y P^{z} z} \nonumber\\ 
&\times \left\langle P+\frac{\Delta}{2} \left| \bar{\psi}\left(-\frac{z}{2}\right) \gamma^{t} U\left(-\frac{z}{2},\frac{z}{2}\right) \psi\left(\frac{z}{2}\right) \right| P-\frac{\Delta}{2} \right\rangle \ ,   
\end{align}
where the quark bilinear operator connected by a gauge link $U$ separates along the spatial $z$ direction, which is directly calculable on the Euclidean lattice.  A handbag diagram of the quasi-GPD is presented in the right panel of Fig.~\ref{fig:handbag}.  The momentum flows along the $z$-direction at the two vertices are $(y+\xi) P^z$ and $(y-\xi) P^z$, respectively. The large-momentum
factorization~\cite{Ji:2015qla,Xiong:2015nua,Liu:2019urm,Ma:2022ggj,Ma:2022gty} takes the form, 
\begin{align}\label{eq:qGPDfac}
\tilde{F}(y,\xi,t,P^z) = \int_{-1}^{1} \frac{d x}{|x|} \tilde{C}\left(\frac{y}{x},\frac{\xi}{x},\frac{\mu}{x P^z}\right) F(x,\xi,t,\mu) \ . 
\end{align}
The advantage of lattice calculations over the virtual Compton process is that the quasi-GPD $\tilde{F}$ can be calculated over the full kinematic region of $y$ and $\xi$.
\begin{figure*}[htbp]
    \centering
    \includegraphics[width=0.40\linewidth]{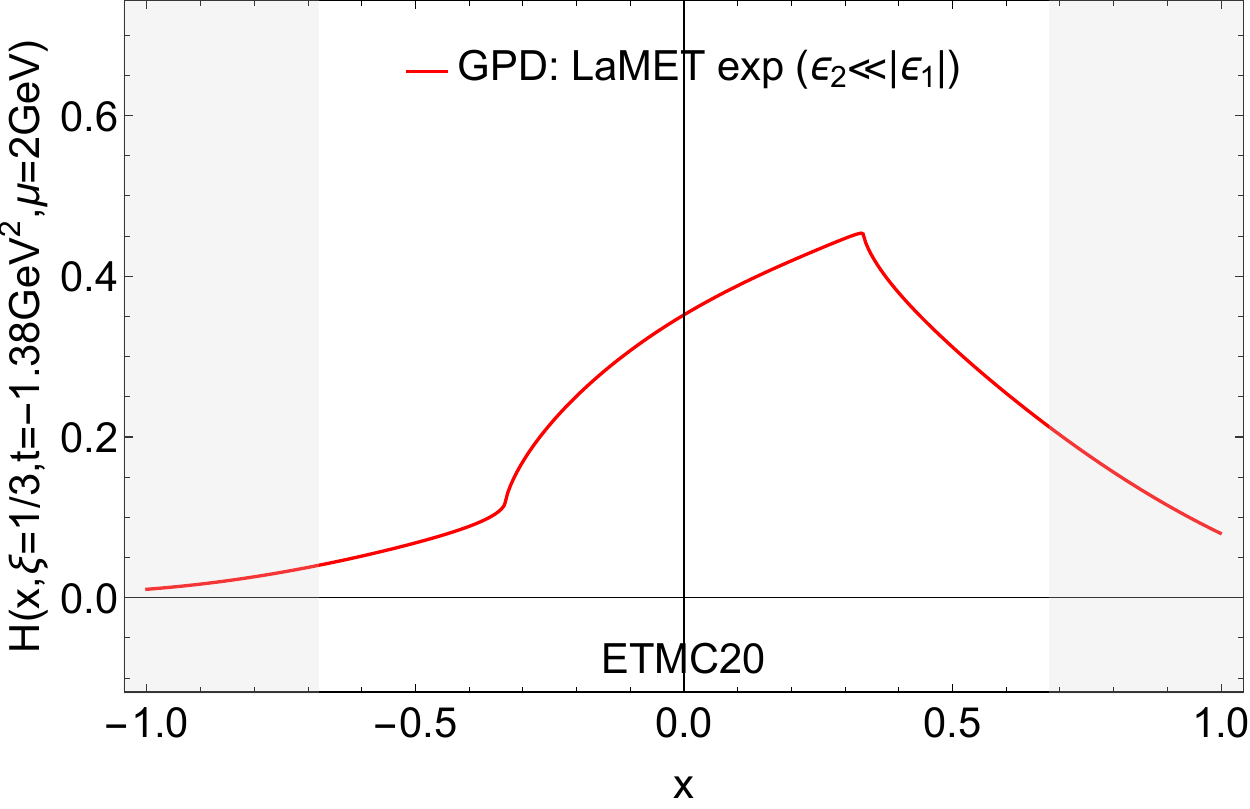}
    \includegraphics[width=0.40\linewidth]{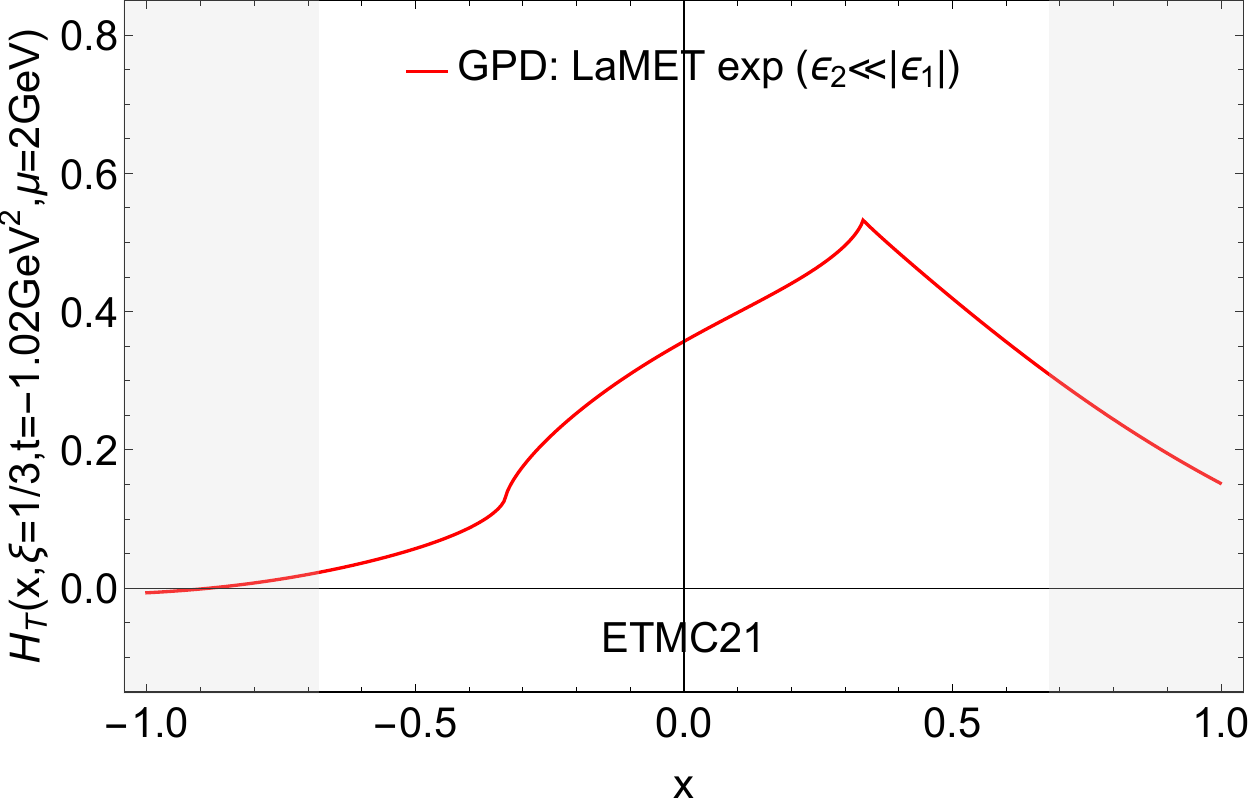}
    \includegraphics[width=0.40\linewidth]{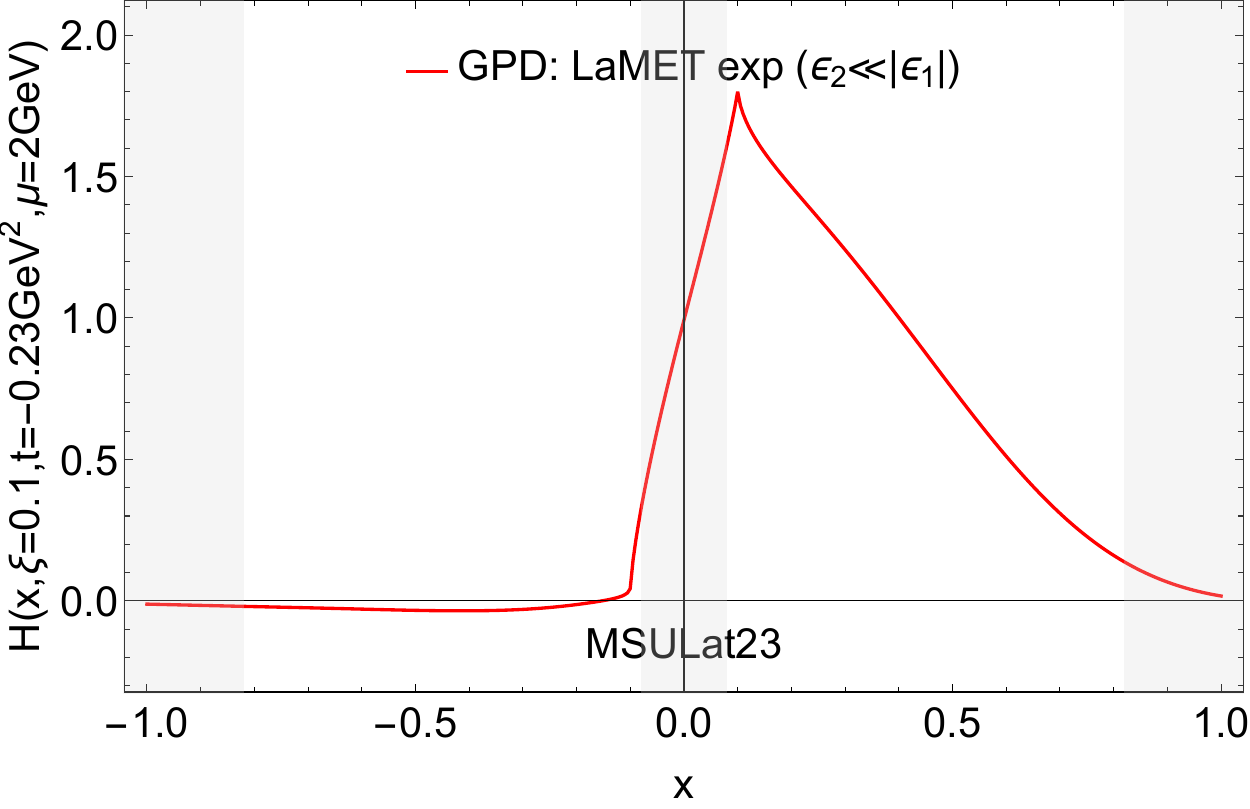}
    \includegraphics[width=0.40\linewidth]{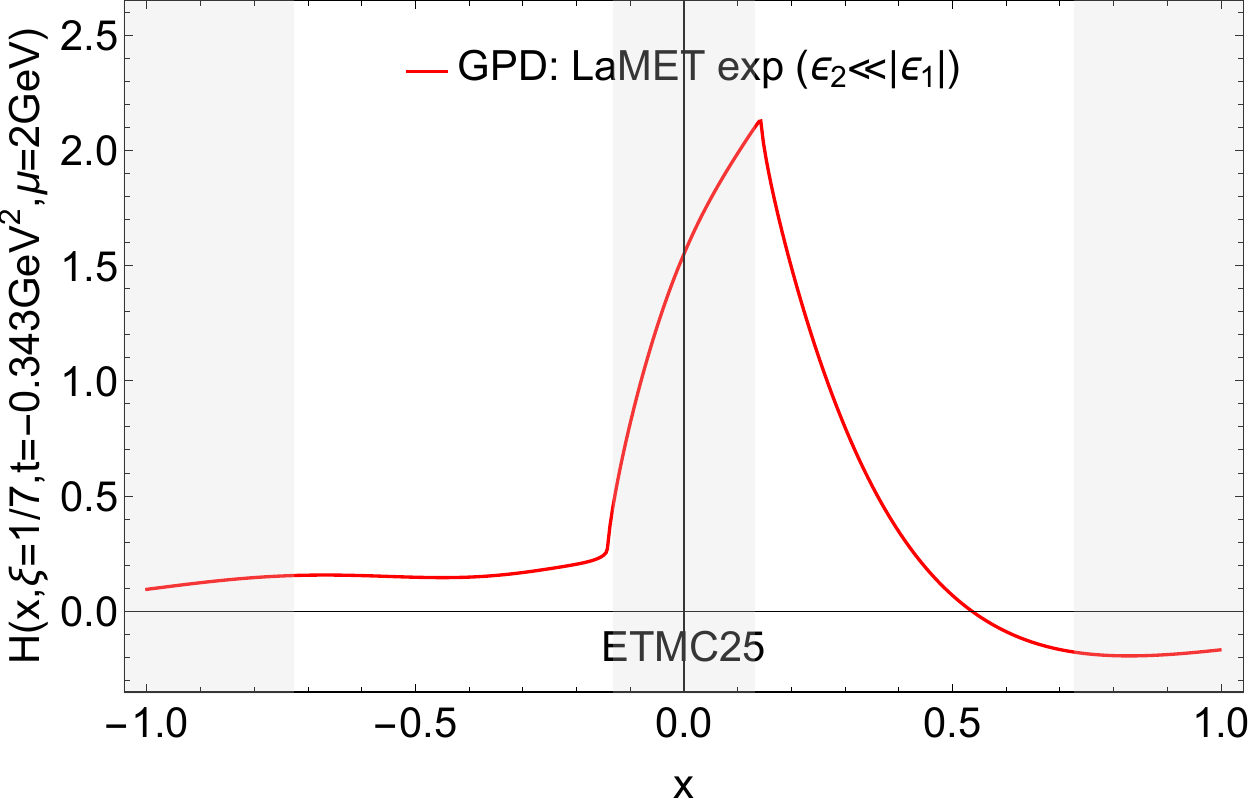}
    \caption{Reanalyzed nonzero-skewness $H$ and $H_T$ GPDs obtained with one-loop matching and the regulator hierarchy $\epsilon_2 \ll |\epsilon_1| \rightarrow 0$. The gray bands indicate regions where either $2|1\pm x|P^z < 0.8 \, {\rm GeV}$ or $2(|x|+|\xi|)P^z < 0.8 \, {\rm GeV}$, for which the large-momentum expansion is not expected to be reliable. }
    \label{fig:HGPD}
\end{figure*}
With that, the quasi-GPD factorization formula Eq.~(\ref{eq:qGPDfac}) can be analytically inverted, yielding a Weinberg-type EFT expansion for light-cone GPD~\cite{Ji:2024oka},
\begin{align}\label{eq:LaMETexp}
F(x,\xi,t,\mu) = \int \frac{d y}{|y|} C\left(\frac{x}{y},\frac{\xi}{y},\frac{\mu}{y P^z}\right) \tilde{F}(y,\xi,t,P^z) \ , 
\end{align}
where the one-loop coefficient $C$ can be found in Ref.~\cite{Ji:2015qla,Xiong:2015nua,Liu:2019urm,Yao:2022vtp}, which contains the following structure sensitive to 
$x\pm\xi,$
\begin{align}\label{eq:qlog}
\left(\frac{1}{2(y-x)}\mp\frac{1}{ 4 \xi }\right) \frac{|x\pm\xi|}{y\pm\xi} \ln \left(\frac{4P_z^2 (x\pm\xi )^2}{\mu ^2}\right) \ ,
\end{align}
plus the associated virtual contributions proportional to $\delta(x/y-1)$. The logarithms related to the scale $2(x\pm \xi)P^z$  are suppressed near $x=\mp \xi$, and there is no leading-twist contribution from the soft scales associated with $(x \pm \xi)P^z$. An illustration of probing GPD at $x=\xi$ with one soft quark is marked in Fig.~\ref{fig:handbag}.

To be sure, the soft quark has infinite light-cone energy in light-front quantization. Such infrared zero-modes disrupt analyticity and cause singular behaviors in light-cone distributions, such as at $x=1$ in the standard parton distribution function (PDF) where there is a derivative discontinuity. Therefore, it is expected that there is a similar derivative discontinuity for GPD at $x=\pm \xi$.

\vspace{0.1in}
\textit{Reanalysis of lattice datasets:}
The pioneering lattice calculations of the nucleon GPDs for $\xi\ne 0$ can be found in Refs.~\cite{Alexandrou:2020zbe,Alexandrou:2021bbo,Holligan:2023jqh,Chu:2025kew}. The relevant information of lattice datasets, including GPD types, skewness, and other parameters, is presented in Tab.~\ref{tab:setlatdat}. It is curious that after one-loop matching, GPD discontinuities have been seen in these works near $x\sim \pm \xi$. If these regions are indeed computable, as we have argued, the observed behavior appears inconsistent with the expectation that GPDs shall be continuous at $x\sim \pm \xi$. 
To resolve this puzzle, we reanalyze the $H$ or $H_T$ datasets. We adopt the large distance asymptotic analysis~\cite{Ji:2020brr,Gao:2021dbh,Chen:2025cxr,Ji:2026vir} to obtain the momentum-space quasi-GPDs $\tilde{F}(y,\xi,t,P^z)$. 

Based on the large momentum expansion in Eq.~(\ref{eq:LaMETexp}), we conduct the perturbative matching up to one-loop accuracy~\cite{Ji:2015qla,Xiong:2015nua,Liu:2019urm,Yao:2022vtp}, 
\begin{align}
C\left(\frac{x}{y},\frac{\xi}{y},\frac{\mu}{y P^z}\right) = \delta\left(1-\frac{x}{y}\right) + \frac{\alpha_s C_F}{2\pi} C^{(1)}\left(\frac{x}{y},\frac{\xi}{y},\frac{\mu}{y P^z}\right) \ .
\end{align}
In this procedure, a careful implementation of regulators is necessary for the reliable prediction of the $x\sim \pm\xi$ region, and our method consists of the following two steps.

First, we calculate the GPD in the $x \sim \pm \xi$ neighborhood, excluding $x = \pm \xi$. The matching kernel at one-loop $C^{(1)}\left(x/y,\xi/y,\mu/yP^z\right)$ is expressed in terms of plus functions, which requires a regulator $\epsilon_2$ near its singular point, 
\begin{align}\label{eq:plus}
&\lim_{\epsilon_2 \rightarrow 0^{+}} \left[\int_{-\infty}^{x-\epsilon_2} + \int^{\infty}_{x+\epsilon_2} \right] dy \nonumber\\
&\times\left[\frac{1}{|y|} C^{(1)}\left(\frac{x}{y},\frac{\xi}{y},\frac{\mu}{y P^z}\right) \tilde{F}(y,\xi,t,P^z) \right. \nonumber\\
&\left. \quad - \frac{1}{|x|} C^{(1)}\left(\frac{y}{x},\frac{\xi}{x},\frac{\mu}{x P^z}\right) \tilde{F}(x,\xi,t,P^z) \right]\ .
\end{align}
After the entire kernel is convoluted with the quasi-GPD, this regulator $\epsilon_2$ can be smoothly taken to zero. This step guarantees the cancellation of the infrared divergence between the real and virtual diagrams. 

Then, since GPD is expected to be continuous at $x=\pm\xi$, its value at $x=\pm\xi$ can be obtained by taking the limit from its neighborhood,
\begin{align}
F(\pm\xi,\xi,t,\mu) = \lim_{\epsilon_1 \rightarrow 0} F(\pm\xi+\epsilon_1,\xi,t,\mu) \ .
\end{align}
Here, $\epsilon_1 \rightarrow 0^{+}$ and $\epsilon_1 \rightarrow 0^{-}$ lead to the same result, which is verified both analytically and numerically with the one-loop matching kernel. That means this step indeed preserves the continuity of GPD. 

The order of limits $\epsilon_2 \rightarrow 0$ and $|\epsilon_1| \rightarrow 0$ does not commute. The above procedure corresponds to $\epsilon_2 \ll |\epsilon_1| \rightarrow 0$, which is well defined. Other limits, including $|\epsilon_1| \ll \epsilon_2 \rightarrow 0$ and $|\epsilon_1| \sim \epsilon_2 \rightarrow 0$, are divergent according to our numerical tests. These observations are understandable because the mathematical properties of a matching kernel, as a generalized function or distribution~\cite{Friedlander:1998}, should be discussed after it is integrated against a smooth test function, and the limit $\epsilon_2 \ll |\epsilon_1| \rightarrow 0$ is consistent with this distributional interpretation.

The final GPD results are shown as red solid curves in Fig.~\ref{fig:HGPD}. The GPDs are continuous at $x=\pm\xi$, although their derivatives are not, which are attributed to the soft-quark (zero-mode) effect as we explained before. This behavior has been seen in GPD models~\cite{Musatov:1999xp,Polyakov:1999gs,Radyushkin:1998es} that preserve the polynomiality condition~\cite{Ji:1998pc,Zhang:2024djl}.
As shown in the figure, the DA-like region smoothly interpolates between the quark and antiquark PDF-like regions, without elaborated structures. Our results provide, for the first time, the first-principles predictions of GPDs in the DA-like region. The apparent discontinuities observed in previous calculations~\cite{Alexandrou:2020zbe,Alexandrou:2021bbo,Holligan:2023jqh,Chu:2025kew} may arise from regulator-related numerical artifacts in the matching procedure. 

\begin{figure}
    \centering
\includegraphics[width=0.8\linewidth]{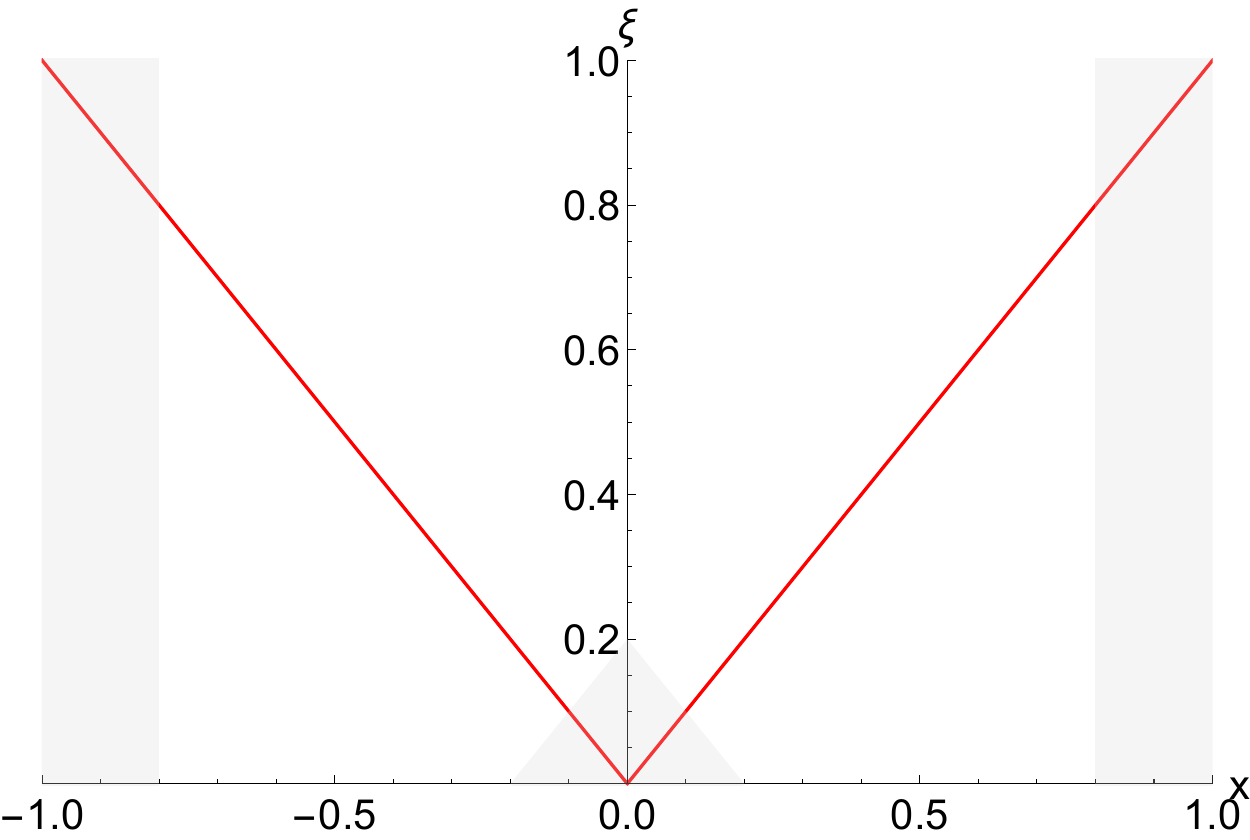}
    \caption{The calculable region of GPDs under LaMET. The unshaded region indicates the kinematics accessible within the large-momentum expansion, where $2(|x|+|\xi|)P^z \gg \Lambda_{\rm QCD}$ and $2|1\pm x| P^z \gg \Lambda_{\rm QCD}$. }
    \label{fig:GPD_region}
\end{figure}

\vspace{0.1in}
\textit{Boundary of computability:} Finally, 
we analyze which regions of GPDs can be reliably accessed under large momentum expansion. The matching kernel~\cite{Ji:2015qla,Xiong:2015nua,Liu:2019urm,Yao:2022vtp} contains the gluon logarithm $\sim \ln(4(x-y)^2 P^2_z/\mu^2)/|x-y|$, whose divergence at $x=y$ is canceled between real and virtual diagrams in Eq.~(\ref{eq:plus}). Near the endpoint regions $x \rightarrow \mp 1$, the phase space of the real integral is suppressed while the virtual integral is not, which causes an incomplete infrared cancelation and effectively introduces the scale $2|1\pm x| P^z$~\cite{Ji:2023pba,Liu:2023onm,Ji:2024hit,Holligan:2025baj}, which should be much larger than $\Lambda_{\rm QCD}$. 

On the other hand, in the region where both $x$ and $\xi$ are small, the scales $2|x+\xi|P^z$ and $2|x-\xi|P^z$ appearing in the quark logarithms in Eq.~(\ref{eq:qlog}) may lead to infrared sensitivity. As we have shown, if one scale is much larger than $\Lambda_{\rm QCD}$ while the other becomes $O(\Lambda_{\rm QCD})$, the contribution associated with the soft scale is power suppressed. However, the LaMET expansion breaks down when both scales become soft. Therefore, the leading-twist LaMET expansion remains applicable provided at least one of the two scales is hard, namely ${\rm Max}\left(2|x+\xi|P^z,2|x-\xi|P^z\right) = 2(|x|+|\xi|)P^z \gg \Lambda_{\rm QCD}$. We summarize the convergent region by the unshaded region in Fig.~\ref{fig:GPD_region}. In particular, the regions $x\sim \pm \xi$ are calculable as long as they are sufficiently separated from $x=0$ and the end-points $x=\pm 1$.

To conclude, we demonstrate that GPDs near $x \sim \pm \xi$ are calculable within the large-momentum expansion. This result is established through theoretical arguments similar to the factorizability of DVCS process involving a final-state real photon. We implement a perturbative matching that ensures infrared cancellation near soft quark boundary and preserves the continuity of GPDs but not their first derivative. This allows a future direct comparison with GPDs extracted from experiments in the regions $x \sim \pm \xi$.

\section*{Acknowledgement}
We thank Minhuan Chu, Krzysztof Cichy, and Yong Zhao for useful discussions and comments. We thank Minhuan Chu for sharing the quasi and light-cone GPDs from Ref.~\cite{Chu:2025kew}. Y. S. is partially supported by the Quark-Gluon Tomography (QGT) collaboration, which is supported by the U.S. Department of Energy (DOE) topical collaboration program (DE-SC0023646). X. J. is partially supported by a grant from National Science Foundation of China, No. 12635004 and Thomas and Linda Lau Family Foundation. Y. L. is supported in part by a Priority Research Area DigiWorld grant under the Strategic Program Excellence Initiative at the Jagiellonian University (Kraków, Poland). R. Z. is supported by the U.S. Department of Energy, Office of Science, Office of Nuclear Physics under grant Contract No. DE-SC0011090 and DOE Quark-Gluon Tomography (QGT) Topical Collaboration under award No. DE-SC0023646.

\bibliographystyle{apsrev4-1}
\bibliography{bibliography}

\end{document}